\documentclass[%
 twocolumn,
 aip,
 apl,
 jmp,%
 amsmath,amssymb,
 reprint,%
]{revtex4-2}

\usepackage{listings}

\usepackage{subfigure}
\usepackage{graphicx}
\graphicspath{{./}{./images/}{./images_all/}{./images_all/spin_circuit_spin_pumping_TI/}}

\usepackage{dcolumn}
\usepackage{bm}

\begin{document}

\preprint{AIP/123-QED}

\title{Spin-circuit representation of spin pumping into topological insulators and determination of giant spin Hall angle and inverse spin Hall voltages}

\author{Kuntal Roy}
\email{kuntal@iiserb.ac.in}
\noaffiliation
\affiliation{Department of Electrical Engineering and Computer Science,\\Indian Institute of Science Education and Research Bhopal, Bhopal, Madhya Pradesh 462066, India}


\begin{abstract}
Topological insulators and giant spin-orbit toque switching of nanomagnets are one of the frontier topics for the development of energy-efficient spintronic devices. Spin-circuit representations involving different materials and phenomena are quite well-established now for its prowess of interpreting experimental results and then designing complex and efficient functional devices. Here, we construct the spin-circuit representation of spin pumping into topological insulators considering both the bulk and surface states with parallel channels, which allows the interpretation of practical experimental results. We show that the high increase in effective spin mixing conductance and inverse spin Hall voltages cannot be explained by the low-conductive bulk states of topological insulators. We determine high spin Hall angle close to the maximum magnitude of one from experimental results and with an eye to design efficient spin devices, we further employ a spin-sink layer in the spin-circuit formalism to increase the effective spin mixing conductance at low thicknesses and double the inverse spin Hall voltage.
\end{abstract}


															
\maketitle

Topological insulators (TIs)~\cite{kosterlitz1972,*kosterlitz1973,*RefWorks:233,*thouless1982,*haldane1983,*haldane1988,kane2005quantum,*kane2005,*bernevig2006HgTe,*fu2007,RefWorks:2734,*RefWorks:2735,*RefWorks:2733,*RefWorks:2631,*kou2017} are exotic materials with intrinsic strong magnetic fields due to giant spin-orbit coupling~\cite{RefWorks:760} causing spin-polarized states with spin-momentum locking. Such remarkable properties of TIs~\cite{RefWorks:2738,RefWorks:2736,RefWorks:2737,RefWorks:1262,he2022topological,RefWorks:2728,RefWorks:2588,RefWorks:2731,RefWorks:2723,RefWorks:2724,RefWorks:2725,RefWorks:2739,bansal2012,lee2016} can harbinger unprecedented technological applications in energy-efficient spintronics~\cite{roy14_3,zutic2004,RefWorks:761,roy20_spin,roy_nanotech_2017x} and quantum computation.~\cite{nayak2008} Electrical measurements also have demonstrated the unique signatures of such spin-polarized states as the measured conductance increases with the TI thickness.~\cite{burkov2010,culcer2010,RefWorks:1276,RefWorks:1260,RefWorks:1269,RefWorks:1271,RefWorks:1272,RefWorks:1273,RefWorks:1270}

A precessing magnet sustained by an externally applied alternating magnetic field~\cite{RefWorks:1168} pumps \emph{pure} spins into the surrounding conductors and the theoretical predictions accord to the experimental results on spin pumping.~\cite{RefWorks:876,roy17_2} According to Onsager's reciprocity,~\cite{RefWorks:1292,*RefWorks:1293} spin pumping and the nonlocal inverse spin Hall effect (ISHE) are the reciprocal phenomena of spin momentum transfer and the direct spin Hall effect (SHE), respectively.~\cite{RefWorks:1295} When the spin pumping contribution is added to the Landau-Lifshitz-Gilbert (LLG) equation~\cite{RefWorks:162,*RefWorks:161} of magnetization dynamics with phenomenological damping parameter,~\cite{RefWorks:647} the experimentally observable quantities are the enhancement of damping and the ferromagnetic resonance phase shift.~\cite{roy20}

Following the footsteps of the giant spin-orbit materials, recent experiments of spin pumping into topological insulators~\cite{RefWorks:2719,RefWorks:1266,RefWorks:1263,RefWorks:1267,RefWorks:1268,RefWorks:1265,RefWorks:2730} have revealed the generation of giant increase in spin mixing conductance and ISHE voltages. Spin pumping mechanism gives us a methodology to estimate the relevant parameters in the system e.g., spin mixing conductance ($g^{\uparrow \downarrow}$) at the magnet-TI interface~\cite{RefWorks:1319,RefWorks:2727}, spin diffusion length ($\lambda$)~\cite{roy17_3,roy_spie_2018x} and spin Hall angle ($\theta_{SH}$), considering the bulk parallel channels between the surface states for practical TIs. Such estimations and understandings can lead to efficient device designs using SHE,~\cite{roy14_3} promising for devising future spintronic devices, alongwith other emerging forerunners.~\cite{roy17,roy16_spin,roy20_spin}

\begin{figure*}
\centering
\includegraphics[width=1\textwidth]{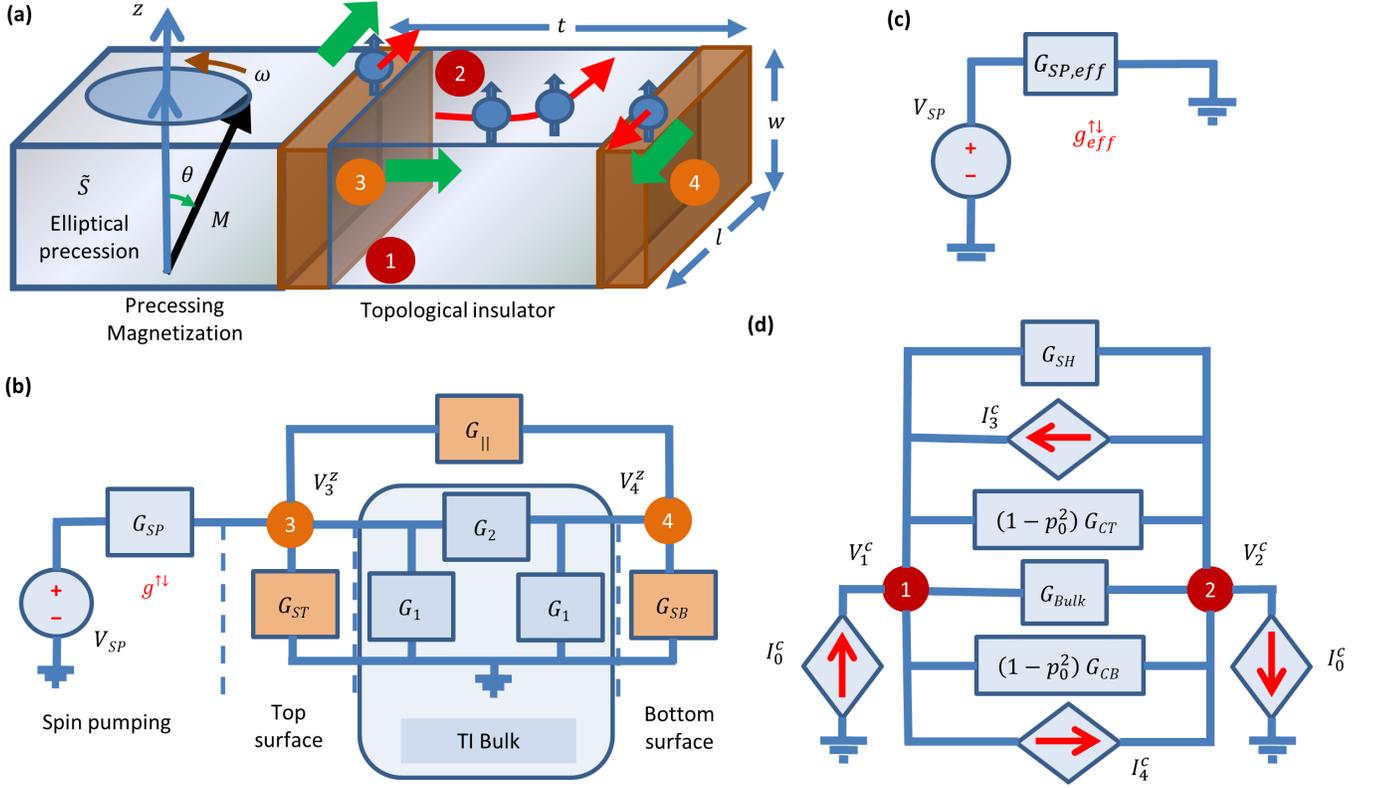}
\caption{\label{fig:spin_circuit_spin_pumping_TI_circuit} (a) A precessing magnetization in a magnetic layer with uniform mode of excitation is pumping a \emph{pure} spin current to the adjacent topological insulator of dimension $l\times w \times t$. With a dc magnetic field along the $z$-direction, and a rf driving field, $\omega$ and $\theta$ are the precession frequency and angle, respectively. Spin potentials are developed on the surfaces marked by 3 and 4, while the charge potentials are generated on the surfaces marked by 1 and 2, due to ISHE. (b) An equivalent spin-circuit representation considering both the bulk states and surface states with parallel channels considering practical devices, with $V_{SP}$ acting as a spin battery, $G_{SP}$ as the interfacial \emph{bare} spin mixing conductance, \emph{two} $\pi$-circuits corresponding to bulk and surface states, as described in text. (c) The \emph{effective} spin mixing conductance $G_{SP,eff}$ of the whole spin-circuit in part (b). (d) The charge-circuit for the spin-to-charge conversion by ISHE with the bulk part (current sources $I_0^c$ and conductance $G_{Bulk}$), top surface (current source $I_3^c$ and conductance $(1-p_0^2)G_{CT}$), bottom surface (current source $I_4^c$ and conductance $(1-p_0^2)G_{CB}$), $p_0$ is the surface polarization, and $G_{SH}$ is the shunting conductance considering the metallic magnet conductance.}
\end{figure*}	

Here, we formulate the spin-circuit representation of spin pumping into topological insulators considering both the TI bulk and surface states with parallel channels for practical TI devices. In spin-circuit formalism, the voltages and currents at different nodes are of 4-components (1 for charge and 3 for spin vector), while the conductances are $4\times 4$ matrices ($c$-$z$-$x$-$y$ basis). Kirchhoff's current and voltage laws have been tremendously successful for the development of the transistor-based technology and there are commercial developments e.g., SPICE (Simulation Program with Integrated Circuit Emphasis).~\cite{hspice} Such spin-circuit theory has been developed for spin pumping, benchmarked with the experimental results, and utilized to propose creative efficient device designs.~\cite{roy17_2,roy17_3,roy_spie_2018x,roy20,roy21_2} The spin-circuit representation of spin pumping in TIs is developed in a systematic manner and simple to follow, which apart from having analytical qualitative picture, can lead to rapid quantitative results of the involved parameter space, particularly for complex structures. We combine it with the spin-circuit model for 2D channels with spin-orbit coupling~\cite{RefWorks:1261} with suitable modifications to incorporate the parallel channels between the TI surface states and benchmark the experimental results in literature.~\cite{RefWorks:2719} It turns out that the high increases in \emph{effective} spin mixing conductance and inverse spin Hall voltages cannot be explained from the low-conductive bulk states, signifying clearly the presence of high-conductive surface states. 

From experimental results,~\cite{RefWorks:2719} the spin Hall angle $\theta_{SH}$ is determined to be 0.75, which is near to the maximum possible magnitude of \emph{one} and very high compared to the giant spin-orbit materials, e.g., Platinum.~\cite{roy17_3} It is pointed out here that if we convert a charge current to spin current and convert it back to the charge current, with Onsager's reciprocity, $\theta_{SH}^2 \leq 1$ (i.e., $|\theta_{SH}| < 1$) to satisfy energy conservation, however, there are many papers (until very recently as well~\cite{RefWorks:2732}) report $\theta_{SH} > 1$. Note that  the gain due to geometric factor considering the dimensions should not be incorporated in the spin Hall angle.~\cite{roy14_3}

We further employ our experimentally benchmarked model to propose and design an efficient device by adding a spin-sink layer adjacent to the bottom surface of a TI, which can prevent degradation of the \emph{effective} spin mixing conductance at low TI thicknesses and hence can \emph{double} the inverse spin Hall voltages. To achieve this, we need to simply write a netlist and solve for different node voltages and currents using a circuit solver. From less complex devices, it may be possible to derive the expressions analytically, however, for more complex devices, any further analytical expression becomes tedious and this depicts the prowess of the spin-circuit approach developed here.

Figure~\ref{fig:spin_circuit_spin_pumping_TI_circuit} shows the spin-circuit representation of spin pumping into topological insulators and also the charge-circuit for the subsequent conversion to charge current due to ISHE. As depicted in the Fig.~\ref{fig:spin_circuit_spin_pumping_TI_circuit}(b), the spin-circuit representation of average spin pumping~\cite{roy17_2} for a complete precession is represented by $V_{SP} = \tilde{S}\,\frac{\hbar \omega}{2e} sin^2\theta$, where $\tilde{S}$ is the frequency dependent elliptical precession factor due to thin magnetic film;~\cite{RefWorks:884} $I_{SP} = lw\,\tilde{S} \, \frac{e \omega}{2\pi}  g_r^{\uparrow \downarrow} sin^2\theta$, thus $G_{SP} = I_{SP}/V_{SP} = lw (2e^2/h) g_r^{\uparrow \downarrow}$, where $g_r^{\uparrow \downarrow}$ is the real part of the \emph{bare} interfacical spin mixing conductance $g^{\uparrow \downarrow}$ (in units of $1/m^2$). Note that first principles calculations and experiments have shown that the imaginary component of $g^{\uparrow \downarrow}$ is negligible for metallic magnets (i.e., $g^{\uparrow \downarrow} \simeq g_r^{\uparrow \downarrow}$).~\cite{RefWorks:1046,RefWorks:814,*RefWorks:813} 

\begin{figure*}
\centering
\includegraphics[width=\textwidth]{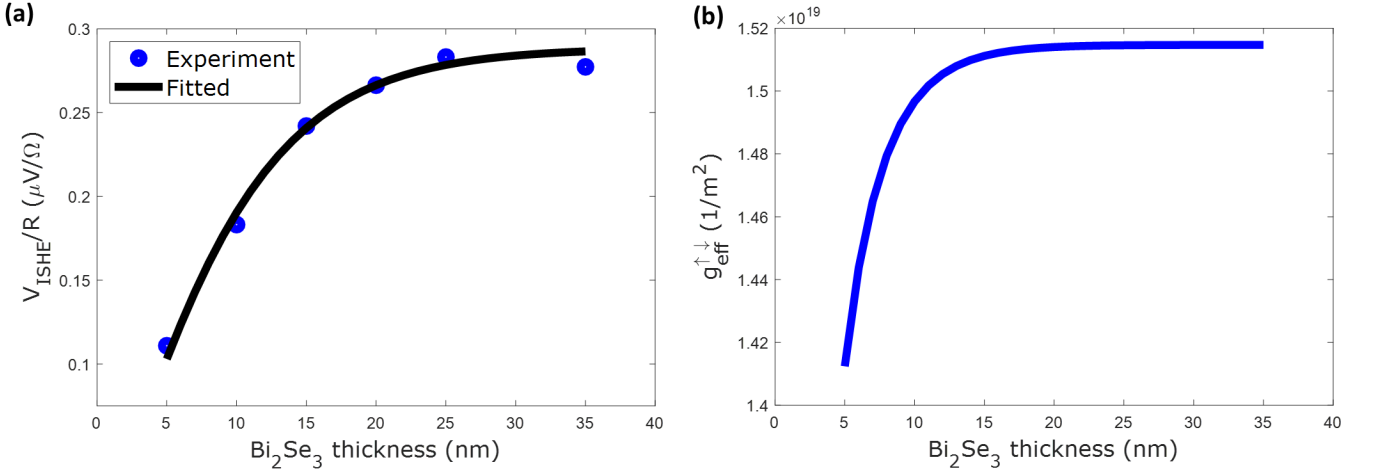}
\caption{\label{fig:spin_circuit_spin_pumping_TI_results} (a) Fitting the thickness dependence (5 -- 35 nm) of inverse spin Hall voltage $V_{ISHE}$ divided by the $Bi_2 Se_3$/$Py$ bilayer resistance $R$, using the spin-circuit theory. The experimental data points are taken from the Ref.~\onlinecite{RefWorks:2719}. (b) The thickness dependence (5 -- 35 nm) of $g_{eff}^{\uparrow \downarrow}$, using the spin-circuit theory.}
\end{figure*}

The TI bulk states are represented by a diffusion based $\pi$-circuit with $G_1 = G_\lambda^b tanh (t/2\lambda_b)$, $G_2 = G_\lambda^b csch (t/\lambda_b)$, $G_\lambda^b = \sigma_b l w/\lambda_b$, where $\sigma_b$ is the bulk conductivity and $\lambda_b$ is the spin diffusion length for the bulk states.~\cite{roy17_2} The spin polarization of the surface states $p_0=(M-N)/(M+N)$, the ballistic conductance $G_B = (e^2/h)(M+N)$, $G_{ST} = G_{SB} = 4 G_B^2/a^2\,G$, $p=a\,p_0$, $M$ and $N$ are the spin-momentum locked number of modes, the ordinary conductance $G = G_B \lambda_{SS}/(\lambda_{SS}+l)$, $\lambda_{SS}$ is the mean free path for the surface states, and the angular averaging factor $a\simeq1$ for TIs.~\cite{RefWorks:1276} Due to the existence of parallel channels between the surface states, we include a conductance $G_{||}$ between the two TI surfaces. We consider a diffusion based model~\cite{roy17_2} for the surface states with parallel channels, and accordingly represent it by a $\pi$-circuit with $G_S = G_{ST} = G_{SB} = G_\lambda tanh (t/2\lambda)$, $G_{||} = G_\lambda csch (t/\lambda)$, $G_\lambda = \sigma l w/\lambda$, where $\sigma$ aand $\lambda$ are the conductivity and spin diffusion length for the surface states, respectively. 

To determine the \emph{effective} spin mixing conductance as depicted in the Fig.~\ref{fig:spin_circuit_spin_pumping_TI_circuit}(c), we can solve the spin-circuit for $V_3^z$ with a circuit solver and using $\left(V_{SP}-V_3^z\right)G_{SP}= V_{SP} \, G_{SP,eff}$, we get 
\begin{equation}
G_{SP,eff}=\left(1-\frac{V_3^z}{V_{SP}}\right) G_{SP}.
\label{eq:G_SP_eff}
\end{equation}

If the conductivity for the surface states is orders more than that of the bulk states, we can consider only the $\pi$-circuit corresponding to the surface states with parallel channels. Accordingly, $G_{TI} = G_S + G_S G_{||}/(G_S + G_{||}) = G_\lambda/coth(t/\lambda)$. Then, $G_{SP,eff} = G_{SP} G_{TI}/(G_{SP} + G_{TI}) = lw (2e^2/h) g_{eff}^{\uparrow \downarrow}$. Hence, we get~\cite{roy17_2}
\begin{equation}
g_{eff}^{\uparrow \downarrow} = \frac{g^{\uparrow \downarrow}}{1+\frac{\lambda}{\sigma}\,\frac{2e^2}{h} g^{\uparrow \downarrow} coth \left(\frac{t}{\lambda}\right) }.
\label{eq:gmix_eff}
\end{equation}
The above equation can be backcalculated to get the \emph{bare} spin mixing conductance $g^{\uparrow \downarrow}$ with the inequality $(2e^2\lambda/h\sigma) g_{eff}^{\uparrow \downarrow} \, coth (t/\lambda) < 1$, as $g^{\uparrow \downarrow} > 0$.~\cite{roy17_2} Note that $g_{eff}^{\uparrow \downarrow}$, not the \emph{bare} $g^{\uparrow \downarrow}$, can be determined from the enhancement of damping in ferromagnetic resonance experiments.~\cite{roy17_2,RefWorks:814,*RefWorks:813}

As depicted in the Fig.~\ref{fig:spin_circuit_spin_pumping_TI_circuit}(d), the charge-circuit represents the generation of inverse spin Hall voltage due to both TI bulk and surface states, $V_{ISHE,TI} = V_2^c - V_1^c$, which also considers the conductance $G_{SH} = \sigma_{m} t_{m} w/l$ due to current shunting through a metallic magnet, where $\sigma_{m}$ and $t_{m}$ are the conductivity and thickness of the magnetic layer, respectively. The charge current sources due to the surface states in Fig.~\ref{fig:spin_circuit_spin_pumping_TI_circuit}(d) are in the opposite directions and are given by $I_{3(4)}^c = (2 p_0 G_B/a) V_{3(4)}^z$ with the conductances $(1-p_0^2) G_{CT(CB)}$, respectively, where $G_{CT(CB)} = G$.~\cite{RefWorks:1276} The charge current due to bulk states is represented by $I_0^c = \beta_b G_{Bulk} \left(V_3^z - V_4^z \right)$, where $G_{Bulk} = \sigma_b t w/l$, $\beta_b = \theta_{SH,b} l/t$, and $\theta_{SH,b}$ is the spin Hall angle corresponding to the bulk states.~\cite{roy17_2}

Applying KCL at node 1 of the charge-circuit in Fig.~\ref{fig:spin_circuit_spin_pumping_TI_circuit}(d), we get $I_0^c = \left(V_1^c - V_2^c \right) \left(G_0 + G_{SH} \right) + \left(I_4^c-I_3^c\right)$, where $G_0 = G_{Bulk} + (1-p_0^2)(G_{CT}+G_{CB})$ and hence
\begin{align}
& V_{ISHE,TI}  \nonumber\\ 
& \hspace*{2mm}= - \left(\frac{2 p_0 G_B}{a G_0} + \beta_b \frac{G_{Bulk}}{G_0}\right) \left(\frac{G_0}{G_0 + G_{SH}}\right) \left(V_3^z - V_4^z \right).
\label{eq:V_ISHE_def}
\end{align}
We here consider $\beta = \theta_{SH} l/t$, where $\theta_{SH}$ is the spin Hall angle corresponding to the \emph{surface} states, similar to the treatment of \emph{bulk} states, rather than the first term in the Equation~\eqref{eq:V_ISHE_def} for practical TI devices and term it as $V_{ISHE}$.

To calculate $\left(V_3^z - V_4^z \right)$, we apply KCL at nodes 3 and 4 of the spin-circuit in Fig.~\ref{fig:spin_circuit_spin_pumping_TI_circuit}(b), and we get
\begin{align}
(V_3^z - V_{SP}) G_{SP} &+ V_3^z (G_1 + G_{ST}) \nonumber\\
& \hspace*{1cm} +  (V_3^z - V_4^z) (G_2 + G_{||}) = 0,
\label{eq:KCL_3}
\end{align}
\begin{equation}
V_4^z (G_1 + G_{SB}) +  (V_4^z - V_3^z)(G_2 + G_{||})= 0.
\label{eq:KCL_4}
\end{equation}
After solving, we get
\begin{align}
V_3^z &=  \frac{G_1 + G_{SB} + G_2 + G_{||}}{D} V_{SP} G_{SP},\\
V_4^z &=  \frac{G_2 + G_{||}}{D} V_{SP} G_{SP},
\label{eq:KCL_sol}
\end{align}
where $D=(G_1 + G_2 + G_{||} + G_{SB})(G_{SP} + G_1 + G_2 + G_{||} + G_{ST}) - (G_2 + G_{||})^2$. 

\begin{figure*}
\centering
\includegraphics[width=1\textwidth]{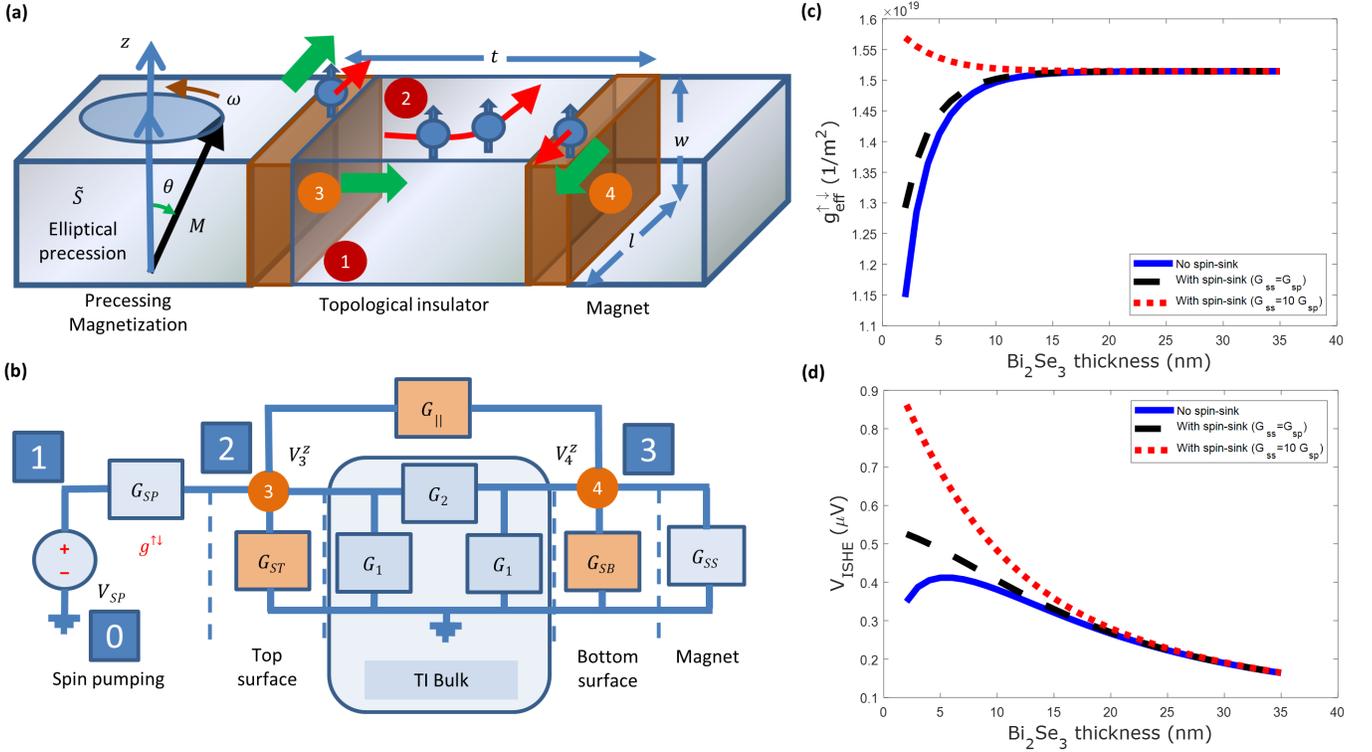}
\caption{\label{fig:spin_circuit_spin_pumping_TI_spin_sink} (a) Proposal of an efficient device design using a spin-sink layer attached to the bottom surface of the TI. (b) The equivalent spin-circuit representation of the proposed design. The node numbers in \emph{squares} are used in the circuit solver. (c) and (d) The thickness dependence (2 -- 35 nm) of (c) $g_{eff}^{\uparrow \downarrow}$ and (d) $V_{ISHE}$, using the spin-circuit theory, when the spin conductance due to the the spin-sink layer $G_{SS}$ is varied 10 times of $G_{SP}$.}
\end{figure*}

Figure~\ref{fig:spin_circuit_spin_pumping_TI_results}(a) shows the benchmarking of the experimental results~\cite{RefWorks:2719} using spin-circuit theory developed here. We use the following equation derived from the spin-circuit theory
\begin{equation}
\frac{V_{ISHE}}{R} = - \theta_{SH}l \lambda e \tilde{S} f  sin^2\theta \; g_{eff}^{\uparrow \downarrow}(t)\, tanh\left(\frac{t}{2\lambda}\right),
\label{eq:V_ISHE}
\end{equation}
where $R$ is the $Bi_2 Se_3$/$Py$ bilayer resistance, $\theta_{SH}$ is the combined spin Hall angle or corresponding to the \emph{surface} states as that dominates over the \emph{bulk} states, FMR frequency $f = \omega/2\pi = 4\;GHz$, and $\theta = 0.23^\circ$.~\cite{RefWorks:2719} The plot in the Fig.~\ref{fig:spin_circuit_spin_pumping_TI_results}(a) is fitted with $\sigma = 50e6\;S/m$, $\lambda = 6.20\;nm$, $\theta_{SH} = 0.75$, and $g^{\uparrow \downarrow} = 1.77e19\;1/m^2$. The elliptical precession factor $\tilde{S}$~\cite{RefWorks:884} is calculated as 0.5 with saturation magnetization $M_s = 1.1\;T$.~\cite{RefWorks:2719} Since the bulk conductivity $\sigma_b = 0.135e6\;S/m$~\cite{RefWorks:2719} is orders lower than the surface conductivity $\sigma = 50e6\;S/m$, it depicts that TI surface states are prominent here. Any thickness dependence of $\lambda$ can be ruled out if the lateral conductance is measured to be thickness-independent.~\cite{bansal2012,roy17_3} The number of modes for the surface states at high TI thicknesses can be determined as $(h/e^2)(\sigma/\lambda) = 2.08e20\;1/m^2$. Note that $g_{eff}^{\uparrow \downarrow}$ has thickness dependence and the plot is given by the Fig.~\ref{fig:spin_circuit_spin_pumping_TI_results}(b), using the spin-circuit theory. Such thickness dependence of $g_{eff}^{\uparrow \downarrow}$ can be determined from experiments and can shed light on the thickness dependence of $\lambda$, if any.~\cite{roy17_3}

Figure~\ref{fig:spin_circuit_spin_pumping_TI_spin_sink} presents an efficient device design using a spin-sink layer with an eye to reduce the degradation of $g_{eff}^{\uparrow \downarrow}$ at low TI thicknesses due to backflow of spins.~\cite{roy17_2,roy17_3} We can simply write a netlist [see the node numbers in squares in the Fig.~\ref{fig:spin_circuit_spin_pumping_TI_spin_sink}(b)] as follows.
\begin{lstlisting}[mathescape,columns=fullflexible,basicstyle=\fontfamily{lmvtt}\selectfont,]
	conductances = 
		[1 2 $G_{SP}$; 2 0 $G_{ST}$; 2 3 $G_{||}$; 3 0 $G_{SB}$; 
		 2 0 $G_1$; 2 3 $G_2$; 3 0 $G_1$; 3 0 $G_{SS}$]
	voltageSources = [1 0 $V_{SP}$]
\end{lstlisting}
With a circuit solver, we can solve for $V_3^z$ and using the Equation~\eqref{eq:G_SP_eff}, we can determine the \emph{effective} spin mixing conductance $g_{eff}^{\uparrow \downarrow}$ of the whole structure, as plotted in the Fig.~\ref{fig:spin_circuit_spin_pumping_TI_spin_sink}(c). The $V_{ISHE}$ can be also calculated using the Equation~\eqref{eq:V_ISHE}, shown in the Fig.~\ref{fig:spin_circuit_spin_pumping_TI_spin_sink}(d). Since the conductivity due to surface states turns out to be high, the conductances of the magnetic layers can be neglected. The results clearly depict that as we increase the spin conductance of the spin-sink layer, the $g_{eff}^{\uparrow \downarrow}$ increases at low thicknesses and it can even increase compared to the high-thickness value. Therefore, the $V_{ISHE}$ keeps increasing at low thicknesses with the addition of the spin-sink layer, otherwise, there is a trade-off in $V_{ISHE}$ with TI thickness \emph{without} a spin-sink layer due to the backflow of spins, which is effectively taken care of by the spin-circuit formalism without any need to incorporate boundary conditions.~\cite{roy17_2} For more complex devices, the analytical expression becomes tedious and this depicts the prowess of the spin-circuit approach.

To summarize, we have developed the spin-circuit representation of spin pumping into topological insulators considering the parallel channels between the surface states. Such treatment allows us to explain the experimental results in literature effectively. The parameters extracted from the model are reasonable and depicts the correctness of the approach utilized here. In particular, the calculations clearly show that the spin Hall angle is near to the maximum magnitude of one. With such experimentally benchmarked model, we have further proposed an efficient design to improve the device characteristics. Such circuits can be simply worked out analytically for qualitative understandings and when more complex they can be solved programmatically to analyze and propose efficient devices.

\vspace*{2mm}
This work was supported by Science and Engineering Research Board (SERB) of India via sanction order MTR/2020/000540.

\vspace*{1mm}
The data that support the findings of this study are available from the corresponding author upon reasonable request.

%

\end{document}